\documentclass[twocolumn,prb,superscriptaddress,showpacs,amsmath,amssymb]{revtex4}
\usepackage{bm,graphicx,dcolumn}
\newcommand{\Hop}{\hat{H}}
\newcommand{\psiop}{\hat{\psi}}
\newcommand{\vecr}{\bm{r}}
\newcommand{\veck}{\bm{k}}

\begin{document}
\title{Intergrain Josephson Currents in Multigap Superconductors: 
Microscopic Origin of Low Intergrain Critical Current and 
Its Recovery Potential in Iron-Pnictide Materials}

\affiliation{
CCSE, Japan Atomic Energy Agency, 
6-9-3 Higashi-Ueno Taito-ku, Tokyo 110-0015, Japan}
\affiliation{
Institute for Materials Research, Tohoku University, 2-1-1 Katahira
Aoba-ku, Sendai 980-8577, Japan} 
\affiliation{
CREST(JST), 4-1-8 Honcho, Kawaguchi, Saitama 332-0012, Japan}
\affiliation{
JST, TRIP, 5 Sanbancho Chiyoda-ku, Tokyo 102-0075, Japan}
\author{Yukihiro Ota}
\affiliation{
CCSE, Japan Atomic Energy Agency, 
6-9-3 Higashi-Ueno Taito-ku, Tokyo 110-0015, Japan}
\affiliation{
CREST(JST), 4-1-8 Honcho, Kawaguchi, Saitama 332-0012, Japan}
\author{Masahiko Machida}
\affiliation{
CCSE, Japan Atomic Energy Agency, 
6-9-3 Higashi-Ueno Taito-ku, Tokyo 110-0015, Japan}
\affiliation{
CREST(JST), 4-1-8 Honcho, Kawaguchi, Saitama 332-0012, Japan}
\affiliation{
JST, TRIP, 5 Sanbancho Chiyoda-ku, Tokyo 102-0075, Japan}
\author{Tomio Koyama}
\affiliation{
Institute for Materials Research, Tohoku University, 
2-1-1 Katahira Aoba-ku, Sendai 980-8577, Japan}
\affiliation{
CREST(JST), 4-1-8 Honcho, Kawaguchi, Saitama 332-0012, Japan}
\date{\today}

\begin{abstract}
We microscopically examine the intergrain Josephson current 
in iron-pnictide superconductors in order to solve the puzzle of 
why the intergrain current is much lower than the intragrain one.
The theory predicts that the intergrain Josephson current 
is significantly reduced by the $\pm s$-wave symmetry when the incoherent
tunneling becomes predominant and the density of states and the gap
amplitude between two bands are identical.   
We find in such a situation that the temperature dependence of the
intergrain Josephson current shows an anomalously flat curve over a wide
temperature range. 
Finally, we suggest important points for increasing the intergrain
current. 
\end{abstract} 

\pacs{74.25.Sv,74.50.+r,74.81.Bd,74.20.Rp,74.81.Fa}
\maketitle

Since the discovery of iron-pnictide
superconductors\cite{Kamihara;Hosono:2008,Takahashi2008}, 
their application potential has been intensively argued. 
It is now widely accepted that they have several advantages
as superconducting wires, tapes, and cables due to high transition
temperature, rich material varieties, and weak anisotropy.  
However, an issue crucial for the transport applications of iron-pnictide
superconductors has been reported by several groups. 
In polycrystalline samples, although the superconductivity is
sufficiently strong inside each grain, the bulk critical current is
unexpectedly small\cite{Yamamoto;Zhao:2008,Tamegai;Eisaki:2008,Wang;Ma:2009,Otabe;Ma:2009,Lee;Larbalestier:2009}.    
This experimental finding indicates that the bulk critical current is
limited by intergrain currents over the grain boundaries in
polycrystalline samples.

In polycrystalline samples of high-$T_{\rm c}$
cuprates\cite{Larbalestier;Polyanskii:2001,Hilgenkamp;Mannhart:2002}
and conventional superconductors\cite{Simanek1994}, the intergrain
coupling at the grain boundaries has been a key topic in assessing their
transport applicability. 
Hence, it is also important in iron-pnictide superconductors to examine
the intergrain coupling and understand their own physics.

In this study, we investigate the superconducting tunneling current at
the grain boundaries in polycrystalline samples of two-band
superconductors described by the two-band BCS Hamiltonian\cite{twoband}. 
We show that the intergrain Josephson current is largely suppressed when
the Cooper pair symmetry is
$\pm s$-wave\cite{Bang;Choi:2008,Mazin;Du:2008,Kuroki;Aoki:2008,Nagai;Machida:2008,Nakai:Machida:2009} 
together with some conditions.   
On the basis of this result, we propose that the bulk critical current
is limited by the reduction mechanism due to the $\pm s$-wave symmetry in
iron-pnictide superconductors. 
The limitation is significant when polycrystalline samples are regarded
as a Josephson-coupled assembly of superconducting grains.  
In addition, we discuss strategies for increasing the intergrain critical
current. 

\begin{figure}[bp]
\centering
\scalebox{0.3}[0.3]{\includegraphics{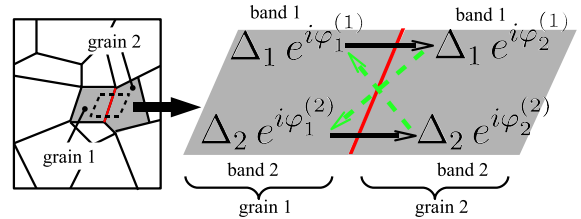}}
\caption{(Color online) Schematic figure of a polycrystalline sample (left-hand side) and focus of an
 intergrain Josephson junction between two-gap superconducting grains (right-hand side), in which  
 two incoherent tunneling channels are schematically depicted. }
\label{fig:grain} 
\end{figure}
 
Throughout this study, we concentrate on a weak-link 
formed between two grains and evaluate its intergrain current, 
assuming the $\pm s$-wave gap symmetry. 
As shown in Fig.\,\ref{fig:grain}, the weak-link is regarded as a 
one-dimensional Josephson junction sandwiched by two superconducting
plates. 
We microscopically calculate the Josephson critical current density
$J_{\rm c}$. 
Such a basic study is necessary prior to examining an ensemble of
weak-links. 

It is known that in Josephson junctions there are two tunneling
processes depending on junction quality. 
One is incoherent
tunneling\cite{Ambegaokar;Baratoff:1963,Koyama:1999,Machida;Tachiki:2000}, 
which does not conserve the momentum of quasi-particles at the tunneling
process.  
This process takes place in junctions with diffusive interfaces. 
The other is coherent
tunneling\cite{Latyshev;Maley:1999,Bulaevskii;Maley:1999,Maki;Haas:2003,Yokoyama;Tanaka:2007},
in which the momentum of the quasi-particles is conserved.  
In most artificially made Josephson junctions, the incoherent tunneling
is predominant, while in high-$T_{\rm c}$ intrinsic Josephson
junctions such as 
${\rm Bi}_{2}{\rm Sr}_{2}{\rm CaCu}_{2}{\rm O}_{8}$
\cite{Latyshev;Maley:1999} 
the coherent tunneling should be considered
as a relevant process in the c-axis transport. 
In this study, we calculate the intergrain Josephson current by taking
account of both the processes.   

In a Josephson junction between multigap (typically two-gap)
superconductors, two channels are possible in the incoherent tunneling,
i.e., intra- and inter-band tunneling channels.  
In the intra (inter)-band tunneling, Cooper pairs can tunnel between
identical (different) bands in the left and right superconductors, as seen in
Fig.\,\ref{fig:grain}, in which the solid (dashed) arrows indicate the
intra (inter)-band Cooper pair tunneling. 
On the other hand, we note that no interband channel exist in the
coherent tunneling.

Let us now investigate the Josephson coupling between the two-gap
superconductors as depicted in Fig.\,\ref{fig:grain}. 
We calculate the Josephson critical current $J_{\rm c}$ in two cases of
the gap symmetry, i.e., $\pm s$-wave and $s$-wave with no sign change, to
show the significant reduction in $J_{\rm c}$ in the $\pm s$-wave case. 
The tunneling Hamiltonian in the present system is written as 
\begin{equation}
\Hop_{{\rm T}}
=
\sum_{a,b,\sigma}
\int {\rm d}^{3}\vecr \int {\rm d}^{3}\vecr^{\prime}
[T^{(ab)}_{\vecr,\vecr^{\prime}}
\psiop^{(a)\dagger}_{\sigma 1}(\vecr)
\psiop^{(b)}_{\sigma 2}(\vecr^{\prime})
+ \text{h.c.} ], 
\end{equation} 
where $\psiop^{(a)}_{\sigma \ell}$ is the field operator for electrons with
spin $\sigma(=\uparrow,\,\downarrow)$ in the $a$-th band in the $\ell$-th
grain ($a=1,2$ and $\ell=1,2$). 
We assume that the tunneling matrix element
$T^{(ab)}_{\vecr,\vecr^{\prime}}$ is real. 
To describe the tunneling processes explicitly, we introduce the
momentum representation of $T^{(ab)}_{\vecr,\vecr^{\prime}}$.  
The Fourier component $T^{(ab)}_{\veck,\veck^{\prime}}$ is
expressed as 
\(
 T^{(ab)}_{\veck,\veck^{\prime}}
= \delta_{ab}\delta_{\veck,\veck^{\prime}}T^{(aa)}_{\veck}
+ T^{\prime (ab)}_{\veck,\veck^{\prime}}
\). 
The first (second) term corresponds to the 
coherent (incoherent) tunneling. 
The incoherent tunneling usually gives a principal contribution in
conventional Josephson
junctions\cite{Ambegaokar;Baratoff:1963,Koyama:1999,Machida;Tachiki:2000}.  
In iron-pnictide polycrystalline systems composed of
randomly oriented grains with rough boundaries, one expects a similar
situation, i.e., the incoherent tunneling is predominant. 
We examine just such a case, in which a tiny coherent
tunneling in addition to a predominant incoherent tunneling exists. 
As shown below, the coherent tunneling alters not only the critical 
current but also its temperature dependence. 

Josephson coupling energy is derived on the basis of the second order
perturbation theory with respect to $\Hop_{\rm T}$. 
In this study, we assume for simplicity that
$(T^{(ab)}_{\veck,\veck^{\prime}})^{2}$ is expressed as
\begin{equation}
 (T^{(ab)}_{\veck,\veck^{\prime}})^{2}
\approx 
\mathcal{T}^{2}[
w\delta_{ab}\delta_{\veck,\veck^{\prime}} + (1-w) ]
\quad (0\le w < 1). 
\end{equation}
In the following, we examine the dependence of intergrain critical current on the
nature of the grain boundary interface, i.e., $w$ (ratio of the coherent
tunneling to incoherent tunneling) dependence. 

On the basis of the standard method employed in several studies\cite{Ambegaokar;Baratoff:1963,Koyama:1999,Machida;Tachiki:2000}, 
the Josephson coupling energy between the two grains is given by 
\begin{equation}
 E_{{\rm J}} 
= -\sum_{a,b}(
\delta_{ab}J^{(aa)}_{{\rm coh}}
+ J^{(ab)}_{{\rm incoh}}
)\cos (\varphi^{(b)}_{2} - \varphi^{(a)}_{1}),
\end{equation} 
where
\begin{eqnarray}
 J^{(aa)}_{{\rm coh}}
&=&
w 
\mathcal{T}^{2}N_{a}^{2} 
3\pi \epsilon_{{\rm F}}L(\beta\Delta_{a}), \label{eq:jj_coh}\\
 J^{(ab)}_{{\rm incoh}}
&=& (1-w) 
\mathcal{T}^{2} N_{a}N_{b} \pi^{2}\Delta^{ab}_{{\rm S}} 
K(k;\beta\Delta^{ab}_{{\rm L}}),
\label{eq:jj_incoh}
\end{eqnarray}
\(
k=
[1-(\Delta^{ab}_{{\rm S}}/\Delta^{ab}_{{\rm L}})^{2}]^{1/2}
\), 
$N_{a}(N_{b})$ is the density of states (DOS) on the Fermi surface for
the electron of the $a(b)$-th  band, $\epsilon_{{\rm F}}$ is the Fermi energy, 
and $\beta(= 1/k_{{\rm B}}T)$ is the inverse temperature. 
The amplitude of the $a(b)$-th order parameter is written as 
$\Delta_{a}(\Delta_{b})(>0)$. 
We define $\Delta^{ab}_{{\rm S}}$ ($\Delta^{ab}_{{\rm L}}$) as 
\(
\Delta^{ab}_{{\rm S}}=\min(\Delta_{a},\,\Delta_{b})
\) 
(\(
\Delta^{ab}_{{\rm L}}=\max(\Delta_{a},\,\Delta_{b})
\)). 
Clearly, 
\(
\Delta^{ab}_{{\rm S}} = \Delta^{ab}_{{\rm L}} = \Delta_{a}
\) when $a=b$.  
The $a$-th superconducting phase in the $\ell$-th grain is expressed as 
$\varphi^{(a)}_{\ell}$. 
The functions $L(\nu)$ and $K(k;\nu)$ are, respectively, defined as
\begin{eqnarray}
L(\nu)
&=& 
\int_{0}^{\pi/2} 
{\rm d}u
\bigg[
\cos u\,
\tanh\bigg(\frac{\nu}{2\cos u}\bigg)
\nonumber \\
&&
\qquad\qquad\qquad
-
\frac{\nu}{2}
{\rm sech}^{2}\bigg(\frac{\nu}{2\cos u}\bigg)
\bigg],\\
K(k;\nu)
&=&
\frac{2}{\pi}
\int_{0}^{1} 
{\rm d}u
\frac{1}{\sqrt{(1-k^{2}u^{2})(1-u^{2})}}
\nonumber \\
&&
\qquad\qquad\qquad
\times 
\tanh\bigg(\frac{\nu\sqrt{1-k^{2}u^{2}}}{2}\bigg). 
\end{eqnarray} 
Here, we note that the difference of chemical potentials between the
grains is assumed to be zero as deriving Eq.\,(\ref{eq:jj_coh}). 

\begin{figure}[tp]
\centering
\scalebox{0.268}[0.268]{\includegraphics{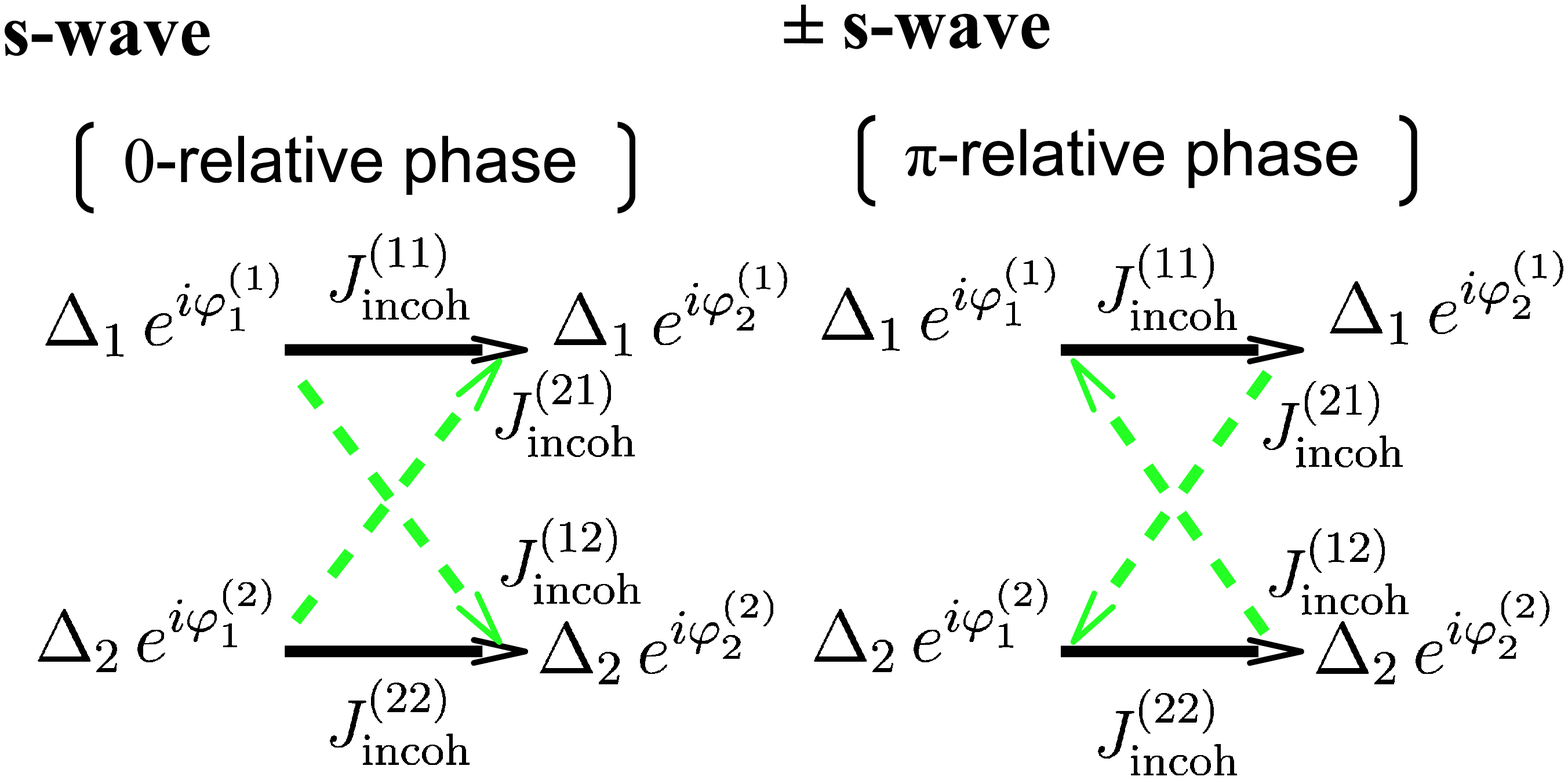}}
\caption{(Color online) Illustration of possible incoherent channels of the intergrain tunneling, in which superposition or cancellation
 occurs depending on the type of the pairing symmetry inside the grains. 
 The symbol
 $J^{(ab)}_{\rm incoh}$ means the Josephson coupling energy corresponding to each
 tunneling channel, and $a,b=1,2$, where $1(2)$ is the band index.}
\label{fig:cancel} 
\end{figure}
Let us evaluate the intergrain critical current from $E_{\rm J}$. 
First, we define the $a$-th band superconducting phase
difference between the two grains as 
\begin{equation}
 \gamma^{(a)} = \varphi^{(a)}_{2} - \varphi^{(a)}_{1}.
\end{equation} 
In term of $\gamma^{(a)}$, we find that
\begin{eqnarray}
\varphi^{(1)}_{2}-\varphi^{(2)}_{1}
&=&\gamma^{(1)} + \chi_{1}, \\
\varphi^{(2)}_{2}-\varphi^{(1)}_{1}
&=&\gamma^{(2)} - \chi_{1},
\end{eqnarray} 
where $\chi_{\ell}$ is the relative phase difference between the two order
parameters in the $\ell$-th grain. 
The pairing symmetry in iron-pnictide superconductors is presently
expected to be the $\pm s$-wave 
symmetry\cite{Bang;Choi:2008,Mazin;Du:2008,Kuroki;Aoki:2008,Nagai;Machida:2008,Nakai:Machida:2009}.
In addition, the recent experiments of the magneto optical
images\cite{Yamamoto;Zhao:2008,Tamegai;Eisaki:2008,Wang;Ma:2009,Otabe;Ma:2009,Kametani;Zhao:2008}
show that superconductivity grows inside each grain. 
Thus, we can naturally assume that $\chi_{\ell}$ should be rigidly fixed
as $\pi$. 
This indicates that  
\(
  \varphi^{(1)}_{2} - \varphi^{(2)}_{1}
= \gamma^{(2)} + \pi
\), 
\(
 \varphi^{(2)}_{2} - \varphi^{(1)}_{1}
= \gamma^{(1)} - \pi
\), and 
\(
 \gamma^{(1)} = \gamma^{(2)} 
\) up to modulo $2\pi$.
Then, we have 
\(
E_{{\rm J}}
=
-J_{{\rm c}}\cos\gamma^{(1)}
\), where 
\begin{equation}
 J_{{\rm c}}
=
\sum_{a}(J^{(aa)}_{{\rm coh}}+J^{(aa)}_{{\rm incoh}})
+ (-1)^{\eta} (J^{(12)}_{{\rm incoh}}+J^{(21)}_{{\rm incoh}}),
\label{eq:ing_jc}
\end{equation}
and $\eta=1$ for the $\pm s$-wave symmetry.
In Eq.\,(\ref{eq:ing_jc}), we find a cancellation mechanism between the currents in the incoherent
tunneling channels, i.e., a cancellation between $J^{(aa)}_{{\rm incoh}}$
and $J^{(ab)}_{{\rm incoh}}$ ($a\neq b$) [Fig.\,\ref{fig:cancel}] for
the $\pm s$-wave symmetry. 
We remark that for the $s$-wave with no sign change, we have $\eta=0$ and no cancellation. 
At $T=0$, we have an explicit formula for $J_{\rm c}(T=0)$ as a function
of $w$, 
\begin{equation}
 J_{{\rm c}}(T=0)
=(1-w) J_{0} \frac{1+(-1)^{\eta}r}{1+r} 
+ w \sigma_{{\rm n}}
\frac{3\hbar \epsilon_{{\rm F}}}{4 e^{2}},
\label{eq:jc_at_zeroT}
\end{equation} 
where 
\begin{eqnarray}
 J_{0}
&=&
\frac{\sigma_{{\rm n}}\pi\hbar}{4e^{2}}
\frac{
N_{1}^{2}\Delta_{1}
+
N_{2}^{2}\Delta_{2}
+
2\kappa N_{1}N_{2}\Delta^{12}_{{\rm S}}}{
(N_{1}+N_{2})^{2}}, \\
\sigma_{{\rm n}}
&=&
\frac{4\pi e^{2}}{\hbar}\mathcal{T}^{2}
(N_{1}+ N_{2})^{2}, \\
r
&=&
\frac{2\kappa N_{1}N_{2} \Delta^{12}_{{\rm S}}}
{(N_{1}^{2}\Delta_{1}+N_{2}^{2}\Delta_{2})},
\end{eqnarray} 
and $e$ is the charge of an electron. 
The positive number $\kappa(\le 1)$ is given by 
$\kappa=2K(k)/\pi$, where $K(k)$ is the complete elliptic integral of
the first kind\cite{Whittaker;Watson:1927}. 
The factor $J_{0}$ is the intergrain critical current for 
$w=0$ (i.e., incoherent tunneling only), and 
$\sigma_{{\rm n}}$ is the normal electron conductivity determined by the
incoherent tunneling. 

\begin{figure}[bp]
\centering
(a)\!\!\!\!\scalebox{0.64}[0.64]{\includegraphics{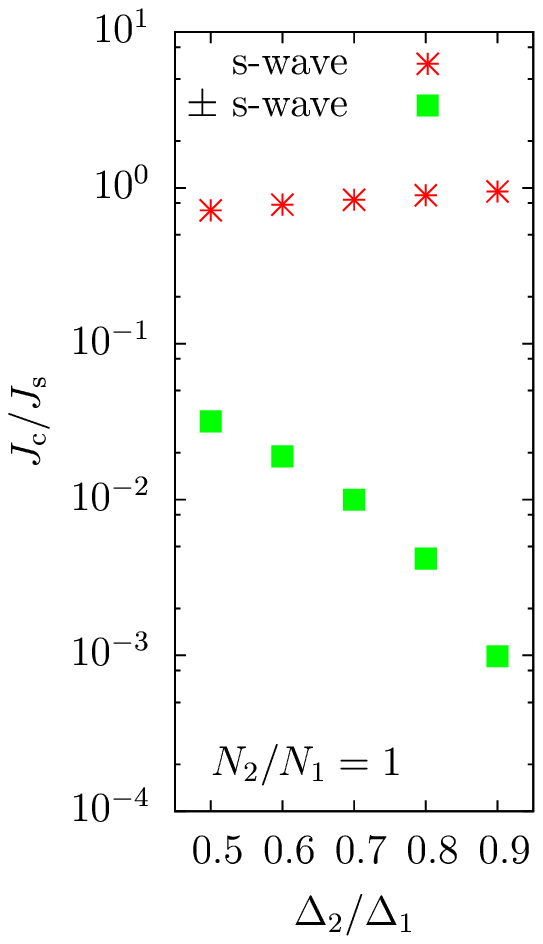}}\quad
(b)\!\!\!\!\scalebox{0.64}[0.64]{\includegraphics{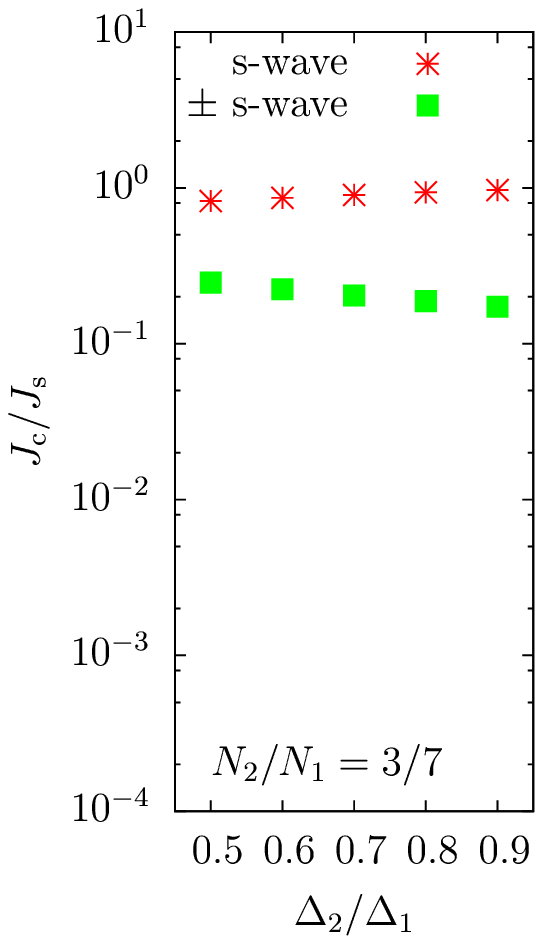}}
\caption{(Color online) Intergrain critical current at $T=0$ in the
 incoherent tunneling only ($w=0$) vs. 
$\Delta_{2}/\Delta_{1}$. 
The unit of the current is
 $J_{{\rm s}}=\sigma_{{\rm n}}\pi\hbar \Delta_{1}/4e^{2}$. 
(a) $N_{2}/N_{1}=1$ and (b) $N_{2}/N_{1}=3/7$.} 
\label{fig:jc_wzero}
\end{figure}

Here, let us evaluate $J_{\rm c}$ in the limiting case, i.e., $w=0$ and
$T=0$, in more detail. 
Figure \ref{fig:jc_wzero} shows that the $\pm s$-wave symmetry leads to
the significant suppression of $J_{{\rm c}}$ compared with the $s$-wave symmetry.
The suppression ratio of the critical current to 
\(
J_{{\rm s}}( =
\sigma_{{\rm n}}\pi \hbar \Delta_{1}(0)/4e^{2} )
\), 
which corresponds to the intergrain $J_{\rm c}$ in the single-gap
$(\Delta_{1})$ superconducting case, strongly depends on the differences
between the DOS and the gap magnitudes of the two bands. 
The theory predicts that the perfect identity of
the superconducting and normal properties between the two different
bands (i.e., $N_{1}=N_{2}$ or $\Delta_{1}=\Delta_{2}$) leads to a great 
suppression for the $\pm s$-wave. 
\begin{figure}[tp]
\centering
(a)\!\!\!\!\!\scalebox{0.56}[0.56]{\includegraphics{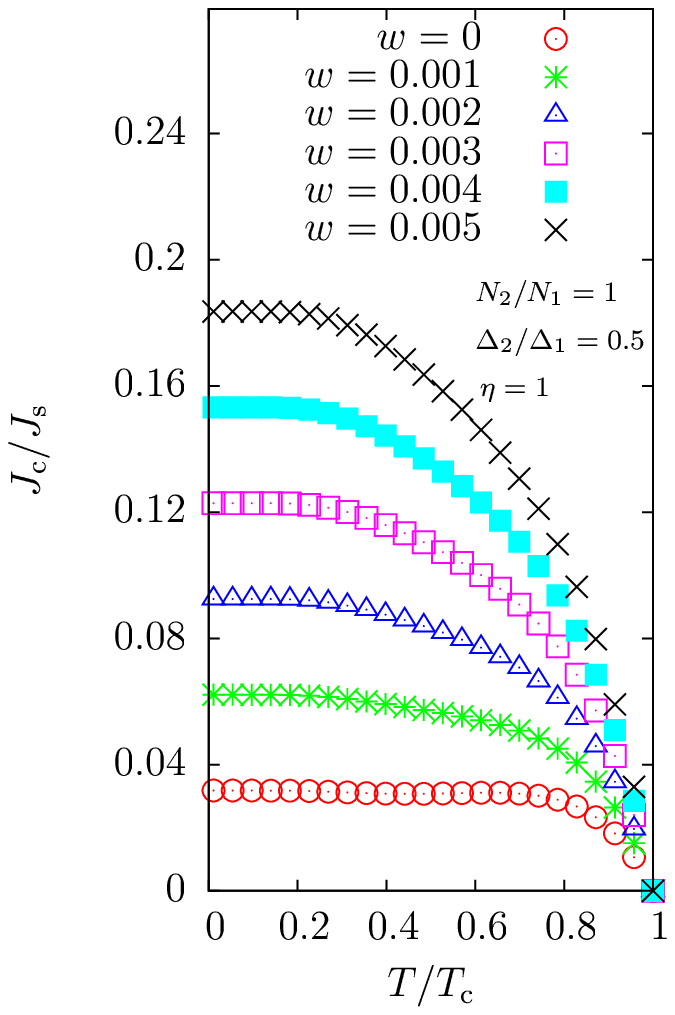}}\quad
(b)\!\!\!\!\!\scalebox{0.56}[0.56]{\includegraphics{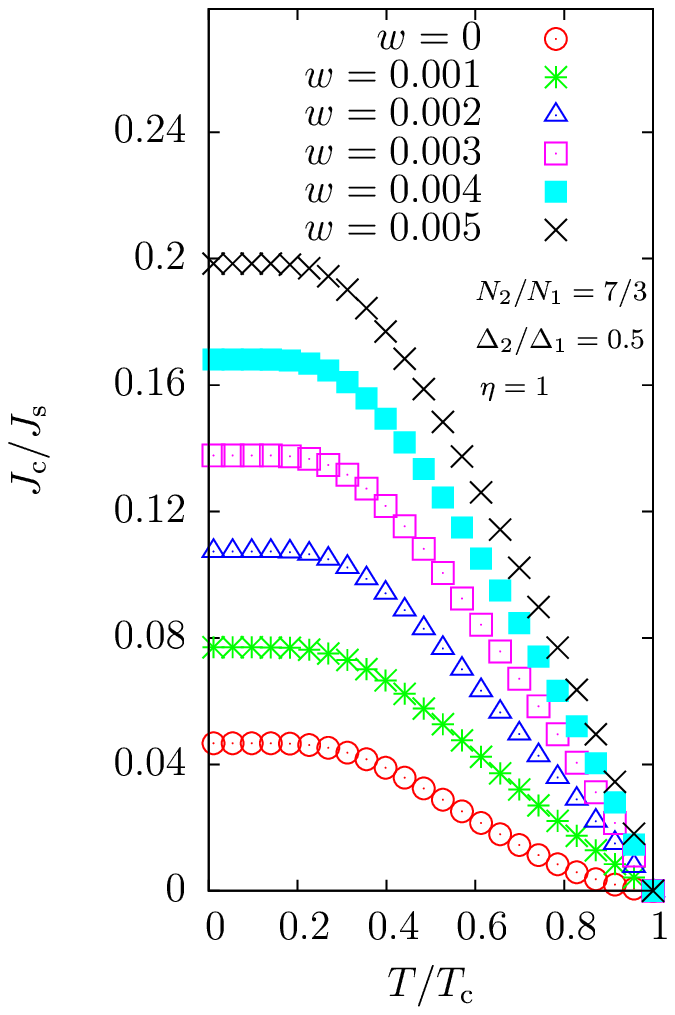}}
\caption{(Color online) Temperature dependence of intergrain critical current with
 coherent-tunneling effect ($w\neq 0$) or without incoherent-tunneling
 effect ($w=0$) for $\pm s$-wave superconductors ($\eta=1$). 
We set $\Delta_{2}/\Delta_{1}=0.5$, 
$\Delta_{1}(0)/\epsilon_{{\rm F}}=10^{-2}$, and   
$\Delta_{1}(0)/k_{{\rm B}}T_{{\rm c}}=3$. 
The unit of $J_{{\rm c}}$ is 
$J_{{\rm s}} = \sigma_{{\rm n}}\pi \hbar \Delta_{1}(0)/4e^{2}$. 
(a) $N_{2}/N_{1}=1$ and (b) $N_{2}/N_{1}=7/3$.} 
\label{fig:jc_td}
\end{figure}

The marked suppression due to the $\pm s$-wave symmetry requires equivalence
between the corresponding two bands. 
However, the equivalence is not perfect in real iron-pnictide
superconductors, since several experiments\cite{Ding;Wang:2008,Gonnelli;Karpinski:2009,Evtushinsky;Borisenko:2009} suggested that
two (in reality more than two) different magnitude gaps coexist in such
materials. 
As for ${\rm Ba}_{0.6}{\rm K}_{0.4}{\rm Fe}_{2}{\rm As}_{2}$, one can
find two gaps in hole-like bands, 
$\Delta_{\alpha}$ and $\Delta_{\beta}$, and one gap in an electron-like
band, $\Delta_{\gamma}$, where 
$|\Delta_{\beta}|\sim |\Delta_{\gamma}|$ and 
$|\Delta_{\beta}|/|\Delta_{\alpha}| \sim 0.5$. 
One can find that  
${\rm Sm}{\rm Fe}{\rm As}{\rm O}_{1-x}{\rm F}_{x}$ has two
different-magnitude gaps, although their relevance to the band structure
remains unclear. 
The value of $N_{2}/N_{1}$ is also not equal to $1$. 
From the first-principles calculation of the electronic
structures\cite{Nakamura}, we can find that, for ${\rm LaFeAs}$
(${\rm BaFe}_{2}{\rm As}_{2}$), the DOS's in the
three hole-like bands are $0.24$, $0.64$, and
$0.49$ ($0.39$, $0.64$, and $0.49$), and the DOS's in the two
electron-like bands are $0.34$ and $0.41$ ($0.42$ and $0.38$). 
Information on the gap amplitudes and band structure is crucial
in evaluating the effect of the $\pm s$-wave symmetry. 
In particular, when the very low intergrain critical 
current density is measured like those in 
\(
{\rm Sm}{\rm Fe}{\rm As}{\rm O}_{1-x}{\rm F}_{x}
\)\cite{Yamamoto;Zhao:2008,Tamegai;Eisaki:2008,Wang;Ma:2009,Otabe;Ma:2009}
and 
\(
{\rm Ba}({\rm Fe}_{1-x}{\rm Co}_{x})_{2}{\rm As}_{2}
\)\cite{Lee;Larbalestier:2009}, 
it is important to examine the suppression ratio induced by the $\pm s$-wave
symmetry.  

In the presence of the coherent tunneling, the above situation is
markedly altered even if the coherent component is tiny. 
From Eq.\,(\ref{eq:jc_at_zeroT}), one understands that the incoherent
tunneling does not become predominant when 
\(
 \Delta_{1},\,\Delta_{2} \ll w\epsilon_{{\rm F}}
\)\cite{Latyshev;Maley:1999}. 
We assume that $\Delta_{1}(T=0)/\epsilon_{{\rm F}}=10^{-2}$ and 
$\Delta_{1}(T=0)>\Delta_{2}(T=0)$ hereafter. 
Figure \ref{fig:jc_td} shows the increase in $J_{{\rm c}}$ by the inclusion of 
the coherent tunneling. 
It is clear that the suppressed $J_{{\rm c}}$ rapidly
recovers by incorporating a small fraction of the coherent tunneling. 
We also find that the temperature dependence of $J_{{\rm c}}$ shows characteristic 
features. 
Here, we note that the temperature dependence of each 
superconducting gap is assumed to obey the weak coupling isotropic
BCS-type theory.  
We use the approximate formula\cite{Carrington;Manzano:2003},  
\(
\Delta_{a}(T) =
\Delta_{a}(0) \tanh\{
1.82[1.018(T_{{\rm c}}/T-1)]^{0.51}\}
\). 
When $w=0$ and $N_{1}=N_{2}$, the temperature
dependence of $J_{\rm c}$ is almost constant over a wide range of
temperature. 
This behavior arises from two facts. 
First, the intra- and inter-band Josephson currents obey almost
equivalent temperature dependences over a wide  
temperature range when $N_{1}=N_{2}$. 
This means that their rates of decrease with respect to temperature 
are almost equal. 
Secondly, the cancellation between them due to the $\pm s$-wave occurs in
Eq.(\ref{eq:ing_jc}). 
Thus, as the two conditions are satisfied, such an anomalous temperature
dependece becomes observable. 
Hence, one can roughly determine the gap symmetry ($\pm s$ or $s$) and the incoherent 
tunneling ratio from the temperature
dependence of $J_{\rm c}$. 

When the incoherent tunneling is predominant, 
the suppression and temperature dependence of $J_{\rm c}$ 
strongly depend on $\Delta_{2}/\Delta_{1}$ and $N_{2}/N_{1}$. 
The suppression rate of $J_{\rm c}$ to $J_{\rm s}$ is $0.004$ for 
$N_{2}/N_{1}=1$ and $\Delta_{2}/\Delta_{1}=0.8$ [Fig.\,\ref{fig:jc_wzero}(a)]. 
When $\Delta_{2}/\Delta_{1}=0.8$, we can find that 
$J_{\rm c}/J_{\rm s}$'s for the $\pm s$-wave are $0.06$, $0.19$, and $0.39$,  
respectively, for $N_{2}/N_{1}=0.67$, $0.42$, and $0.25$. 
The flat temperature dependence of $J_{\rm c}$ in
Fig.\,\ref{fig:jc_td}(a) is observed when $N_{1}\sim N_{2}$.

Finally, let us discuss important points to enhance the intergrain
critical current. 
As shown above, the coherent tunneling induces a drastic enhancement of
$J_{\rm c}$.  
We emphasize that even if the percentage of the coherent component
in the tunneling process is only 0.5\%, $J_{\rm c}$ becomes about
fivefold larger. 
This can be achieved by improving the weak-link (junction)
properties\cite{Wang;Qi;Ma:2009}.  
On the other hand, in the case where no coherent tunneling exists the
suppression of $J_{\rm c}$ strongly depends on the 
physical equivalence between the different bands. 
We have found that quite a large suppression in $J_{\rm c}$ occurs when
$\Delta_1=\Delta_2$ and $N_1=N_2$ in the $\pm s$-wave case. 
Moreover, we point out that the suppression is not so sensitive to the
misorientation angle between neightboring grains due to the full gap
compared to the angle-sensitive cuprate-superconductors\cite{Lee;Larbalestier:2009,Larbalestier;Polyanskii:2001,note1}.
This consideration simply indicates that materials with $\Delta_1 \neq \Delta_2$ and 
$N_{1}\neq N_{2}$ are preferable in avoiding the large suppression of $J_{\rm c}$.
Of course, $J_{\rm c}$ in real polycrystalline samples depends on the
global network structure between grains\cite{Wang;Qi;Ma:2009}, and an
ultimate solution for solving the weak-link problem 
at grain boundaries is to grow a quasi-single  
crystal as large as possible\cite{Larbalestier;Polyanskii:2001}.  

In summary we investigated the intergrain Josephson current in
polycrystalline two-band superconductors with the $\pm s$-wave symmetry
by microscopically deriving the Josephson coupling energy. 
The present theory reveals that the intergrain Josephson current is
significantly reduced by the existence of inter-band tunneling
channels when the incoherent tunneling is predominant. 
We also found the anomalous temperature dependence in the limit at which
the incoherent tunneling is predominant.
Finally, we discussed the important points for increasing the intergrain
Josephson current. 

\begin{acknowledgments}
The authors (Y.O. and M.M) wish to acknowledge valuable discussions with
T. Tamegai, Y. Takano, E. S. Otabe, S. Awaji, N. Hayashi, Y. Nagai,
S. Shamoto, H. Nakamura, M. Okumura, and N. Nakai. 
The work was partially supported by a Grant-in-Aid for Scientific Research
on Priority Area ``Physics of new quantum phases in superclean
materials'' (Grant No. 20029019) from the Ministry of Education,
Culture, Sports, Science and Technology of Japan. 
\end{acknowledgments}

\end{document}